\documentclass[12pt]{article}
\usepackage{graphics}
\usepackage{color}
\providecommand{\SU}[1]{}
\renewcommand{\SU}[1]{\ensuremath{\mathrm{SU}(#1)}}
\providecommand{\UI}{}
\renewcommand{\UI}{\ensuremath{\mathrm{U}(1)}}
\providecommand{\psgroup}{}
\renewcommand{\psgroup}{\ensuremath{\mathrm{SU}(4) \otimes \mathrm{SU}(2)_L \otimes \mathrm{SU}(2)_R } }
\providecommand{\smgroup}{}
\renewcommand{\smgroup}{\ensuremath{\mathrm{SU}(3) \otimes \mathrm{SU}(2)_L \otimes \mathrm{U}(1)_Y } }

\providecommand{\Dpbranes}{}
\renewcommand{\Dpbranes}{\ensuremath{Dp\mathrm{-branes}}}
\providecommand{\Dbrane}[1]{}
\renewcommand{\Dbrane}[1]{\ensuremath{D#1\mathrm{-brane}}}
\providecommand{\Dbranes}[1]{}
\renewcommand{\Dbranes}[1]{\ensuremath{D#1\mathrm{-branes}}}
\providecommand{\vev}{}
\renewcommand{\vev}{vev}
\providecommand{\vevs}{}
\renewcommand{\vevs}{vevs}

\begin{document}
\baselineskip 24pt

\newcommand{\sheptitle}
{Lepton Flavour Violation from Yukawa Operators, 
Supergravity and the See-Saw Mechanism}

\newcommand{\shepauthor}
{S. F. King and I. N. R. Peddie}

\newcommand{\shepaddress}
{Department of Physics and Astronomy, University of Southampton \\
        Southampton, SO17 1BJ, U.K}
\newcommand{\shepabstract}
{We investigate the phenomenological impact of different sources
of lepton flavour violation arising from realistic
models based on supergravity mediated supersymmetry breaking
with Yukawa operators.
We discuss four distinct sources of lepton flavour violation
in such models: minimum flavour violation, arising from neutrino
masses and the see-saw mechanism with renormalisation group (RG)
running; supergravity flavour violation due to the non-universal
structure of the supergravity model; flavour violation due to 
Froggatt-Nielsen (FN) fields appearing in Yukawa operators
developing supersymmetry breaking F-terms and
contributing in an non-universal way to soft trilinear terms;
and finally heavy Higgs flavour violation arising from the heavy 
Higgs fields used to break the unified gauge symmetry which
also appear in Yukawa operators and
behave analagously to the FN fields. In order to quantify the relative
effects, we study a particular type I string inspired model based
on a supersymmetric Pati-Salam model arising from intersecting
D-branes, supplemented by a $U(1)$ family symmetry.}

\begin{titlepage}
\begin{flushright}
  hep-ph/0307091 \\
  SHEP/0320
\end{flushright}
\begin{center}
  {\large{\bf \sheptitle}}
  \\ \shepauthor \\ \mbox{} \\ {\it \shepaddress} \\
  {\bf Abstract} \\ \bigskip 
\end{center}
\setcounter{page}{0} 
\shepabstract
\begin{flushleft}
  \today
\end{flushleft}
\end{titlepage}

\newpage 
\section{Introduction} 
\label{sec:introduction}

Lepton flavour violation (LFV) has been long known to be a sensitive
probe of new physics in supersymmetric (SUSY) models 
\cite{Borzumati:1986qx,Gabbiani:1996hi}.
LFV arises in SUSY models due to off-diagonal slepton masses
in the super-CKM basis in which the Yukawa matrices are diagonal.
Such flavour violation could arise either directly at the high
energy scale due to primordial string or SUGRA effects, or
be generated radiatively by the renormalisation group equations,
for example in running a grand unified theory (GUT) from the
Planck scale to the GUT scale (due to the presence of Higgs triplets)
\cite{Barbieri:1995tw}
or in running the minimal supersymmetric standard model (MSSM) 
with right-handed neutrinos from the Planck scale down to 
low energies, through the scales at which the right-handed neutrinos
decouple. 

Even in
minimal supergravity (mSUGRA), where there is no flavour violation
at the high energy scale, the presence of heavy right-handed neutrinos
as required by the see-saw mechanism explanation of small neutrino 
masses will lead inevitably to LFV \cite{Hisano:1995cp,King:1998nv}.
The recent neutrino experiments which confirm the matter enhanced
Large Mixing Angle (LMA) solution to the solar neutrino problem
\cite{GoPe}, together with the atmospheric data
\cite{SKamiokandeColl}, show that neutrino masses are inevitable,
and, assuming SUSY and the see-saw mechanism, hence show that LFV
must be present. For example this has recently been studied in mSUGRA 
models with a natural neutrino mass hierarchy \cite{Blazek:2002wq}.
There is in fact a large literature on this subject \cite{huge}. 

Recently it has been realised that
in realistic string inspired models based on supergravity mediated 
supersymmetry breaking, in which the origin of Yukawa matrices $Y_{ij}$
is due to Froggatt-Nielsen (FN) operators \cite{Froggatt:1978nt}
of the form 
$Y_{ij}\sim \theta^{n_{ij}}$, where $n_{ij}$ is an integer power,  
there may be a new and dangerous
source of LFV which arises when the FN fields $\theta$ develop
supersymmetry breaking F-terms $F_{\theta}\sim m_{3/2}\theta$ 
leading to non-universal soft trilinear terms 
$\Delta A_{ij}=F_{\theta}\partial_{\theta}\ln Y_{ij}$ \cite{Dudas:1995eq}
which implies $\Delta A_{ij}\sim n_{ij}m_{3/2}$ 
\cite{Abel:2001ur,Ross:2002mr}.
The effect is independent of the vacuum expectation value (vev)
of the FN field $\theta$, and is present even in minimum flavour 
violation scenarios such as mSUGRA. 

In this paper we shall explore the phenomenological impact
of the new source of LFV arising from FN fields discussed above,
and compare it to the more usual sources of LFV arising from
right-handed neutrinos, and non-universal SUGRA models in order
to gauge its relative importance. 
A phenomenological analysis is necessarily model dependent,
and so we shall study a particular type I string inspired model based
on a supersymmetric (SUSY) Pati-Salam model arising from intersecting
D-branes, which was introduced in \cite{Everett:2002pm}.
However in order to explore the effects of interest, it is necessary
to supplement this model by a $U(1)$ family symmetry,
and introduce FN fields so as to provide a realistic description
of quark and lepton masses and mixing angles, including those
of the neutrino sector. Recently a global $\chi^2$ analysis of
a realistic SUSY Pati-Salam model was performed \cite{Blazek:2003wz},
and a good fit to the quark and lepton mass spectrum was obtained
based on a FN operator analysis with a $U(1)$ family symmetry.
It is therefore natural to combine the models in \cite{Everett:2002pm}
and \cite{Blazek:2003wz} in order to provide a realistic 
framework for studying the new LFV effects arising from the FN fields,
and to compare this to the effects on non-universal SUGRA and also
right-handed neutrinos in a model that gives a good fit to the
neutrino data. 

Of course in combining the two models we are 
taking some liberties with string theory. In particular
we assume that the combined model corresponds to the low energy
limit of a string model as in \cite{Everett:2002pm}, but 
with the addition of an extra state, 
which is a Froggatt-Nielsen \cite{Froggatt:1978nt} family
field, $\theta$. We assume that since the model without $\theta$ can
be extracted from a string model, then so can the model with $\theta$,
but we make no attempt to derive it. We emphasise that the main
motivation for combining the two approaches is to explore the
phenomenology of LFV in a ``realistic'' framework. One by-product
of doing this is that we identify a genuinely new source of 
LFV that has not been considered at all in the literature,
namely the heavy Higgs fields $H$ that break the unified gauge symmetry
at high energies. These heavy Higgs fields also appear in the
operators which describe the Yukawa couplings, and they can be
expected to behave in a similar way to the FN fields $\theta$,
and give rise to LFV analagously.
The combined model has a number of attractive features: 
it includes approximate third family
Yukawa unification, the number of free parameters is restricted 
to eight undetermined free parameters related to 
supergravity, and the model gives a good fit to all quark and
lepton masses and mixing angles.

In order to study the phenomenological
effect of the different sources of LFV, we generalise the Goldstino
Angle parametrisation of the dilaton and moduli fields $S,T_i$
to include a parametrisation
of the SUSY breaking F-terms for the FN fields $\theta$ 
and heavy Higgs fields $H$.
There are four distinct sources of lepton flavour violation
in this model: minimum flavour violation, arising from neutrino
masses and the see-saw mechanism with renormalisation group (RG)
running; supergravity flavour violation due to the non-universal
structure of the supergravity model; FN flavour violation due to 
the FN fields developing supersymmetry breaking F-terms and
contributing in an non-universal way to soft trilinear terms;
and finally heavy Higgs flavour violation arising from the 
Higgs fields used to break the unified gauge symmetry which may
behave analagously to the FN fields. We propose four
benchmark points at which each of these four sources separately dominate.
We then perform a detailed numerical
analysis of LFV arising from the four benchmark
points. The numerical results show that LFV due to FN fields is the
most sensitive source in the sense of leading to 
larger limits of $m_{3/2}$, however we find that the
gluino mass is relatively light in these cases which tends
to reduce fine-tuning. 
We also find that in some cases the LFV effects from Yukawa operators
in the presence of the seesaw mechanism can be less than without the
seesaw mechanism.

The outline of this paper is as follows. In section 
\ref{sec:442-pati-salam} we
introduce the specific model that we shall study,
discuss the symmetries of the model, and the Yukawa and Majorana
operators, and for particular choices of the order unity coefficients,
show that this leads to a good fit to the neutrino
data, with a prediction for the unmeasured $\theta_{13}$.
In section \ref{soft} we discuss the soft SUSY breaking
aspects of the model. We parametrise the SUSY breaking F-terms,
give the soft scalar masses, including the D-term contributions,
give the soft gaugino masses and soft tirlinear masses, and explain
why these are expected to lead to large flavour violation.
In section \ref{sec:numerical-etc} we give the results of a 
numerical analysis of the model, focussing on four benchmark
points designed to highlight the four different sources of LFV.
Finally we present our concluding remarks in section \ref{sec:conclusions}

\section{The Model}
\label{sec:442-pati-salam}

\subsection{Symmetries and Symmetry Breaking}
The model defined in Table~\ref{tab:particle_content_42241}
is an extention of the Supersymmetric
Pati-Salam model discussed 
in ref.\cite{Everett:2002pm}, based on two $D5$ branes which intersect
at $90$ degrees and preserve SUSY down to the TeV energy scale.
The string scale is taken to be equal to the GUT scale, about
$3\times 10^{16}$ GeV. 

\begin{table}[htbp]
  \centering \framebox[16cm]{
  \begin{tabular}{|c||c|c|c|c||c|c|c|}
    \hline Field & $\mathrm{SU}(4)^{(1)}$ & $\mathrm{SU}(2)_L$ &
    $\mathrm{SU}(2)_R$ & $\mathrm{SU}(4)^{(2)}$ & Ends & $\UI_F$
    charge & $\UI_{\overline{F}}$ charge \\ \hline $h$ & 1 & 2 & 2 & 1
    & $C_1^{5_1}$ & $0$ & $0$ \\ $F_3$ & 4 & 2 & 1 & 1 & $C_2^{5_1}$ &
    $\frac{5}{6}$ & $0$ \\ $\overline{F}_3$ & $\overline{4}$ & 1 & 2 &
    1 & $C_3^{5_1}$ & $\!\!\!\!\!-\frac{5}{6}$ & $0$ \\ $F_2$ & 1 & 2
    & 1 & 4 & $C^{5_1 5_2}$ & $\frac{5}{6}$ & $2$ \\ $\overline{F}_2$
    & 1 & 1 & 2 & $\overline{4}$ & $C^{5_1 5_2}$ & $\frac{7}{6}$ & $0$
    \\ $F_1$ & 1 & 2 & 1 & 4 & $C^{5_1 5_2}$ & $\frac{11}{6}$ & $1$ \\
    $\overline{F}_1$ & 1 & 1 & 2 & $\overline{4}$ & $C^{5_1 5_2}$ &
    $\frac{19}{6}$ & $4$ \\ $H$ & 4 & 1 & 2 & 1 & $C_1^{5_1}$ &
    $\frac{5}{6}$ & $0$ \\ $\overline{H}$ & $\overline{4}$ & 1 & 2 & 1
    & $C_2^{5_1}$ & $\!\!\!\!\!-\frac{5}{6}$ & $0$ \\ $\varphi_1$ & 4
    & 1 & 1 & $\overline{4}$ & $C^{5_1 5_2}$ & $-$ & $-$ \\
    $\varphi_2$ & $\overline{4}$ & 1 & 1 & 4 & $C^{5_1 5_2}$ & $-$ &
    $-$ \\ $D_6^{(+)}$ & 6 & 1 & 1 & 1 & $C_1^{5_1}$ & $-$ & $-$ \\
    $D_6^{(-)}$ & 6 & 1 & 1 & 1 & $C_2^{5_2}$ & $-$ & $-$ \\ $\theta$
    & 1 & 1 & 1 & 1 & $C^{5_1 5_2}$ & $-1$ & $-1$ \\
    $\overline{\theta}$ & 1 & 1 & 1 & 1 & $C^{5_1 5_2}$ & $1$ & $1$ \\
    \hline
  \end{tabular}
}
  \caption{The particle content of the 42241 model, and the brane
    assignments of the corresponding string}
  \label{tab:particle_content_42241}
\end{table}

The extension is to include an additional $U(1)_F$ family symmetry and the 
FN operators as in \cite{Blazek:2003wz}
(see also \cite{King:OperatorAnalysis}).
The present
42241 Model is then just the 4224 Model of \cite{Everett:2002pm}
augmented by a $\UI_F$ family symmetry. 
The purpose of this extension is to allow a more realistic
texture in the Yukawa trilinears $Y_{abc}$, along the lines
of the recent operator analysis in \cite{Blazek:2003wz}.

The quark and lepton fields are contained in the representaions 
$F, \overline{F}$ which are assigned
charges $X_F$ under $\UI_F$. In Table~\ref{tab:particle_content_42241}
we list two equivalent sets of charges $\UI_F$ and
$\UI_{\overline{F}}$, where $\UI_F$ is anomaly free,
but $\UI_{\overline{F}}$ is equivalent for all practical purposes
and has much simpler charge assignments.
The field $h$ represents both Electroweak Higgs doublets that
we are familiar with from the MSSM.  The fields $H$ and $\overline{H}$
are the Pati-Salam Higgs scalars;
\footnote{We will also refer to these as ``Heavy Higgs''; this
has nothing to do with the MSSM heavy neutral higgs state $H^0$} the bar on the second is used to
note that it is in the conjugate representation compared to the
unbarred field.

The extra Abelian $U(1)_F$ gauge group is a family symmetry, and is broken at
the high energy scale by the \vevs{} of the FN fields \cite{Froggatt:1978nt} 
$\theta,\overline{\theta}$, which have charges $-1$ and $+1$
under $U(1)_F$, respectively. 
We assume that the singlet fields $\theta,\overline{\theta}$ 
arise as intersection
states between the two \Dbranes{5}, transforming under the remnant
\UI s in the 4224 gauge structure. In general they are expected to 
have non-zero F-term \vev s.

The two $SU(4)$ gauge groups are broken to their diagonal
subgroup at a high scale due to the assumed \vevs{} of the 
fields $\varphi_1$, $\varphi_2$ \cite{Everett:2002pm}.
The symmetry breaking at the scale $M_X$
\begin{equation}
  \label{eq:gauge_breaking_pattern}
  \psgroup \rightarrow \smgroup
\end{equation}
is achieved by the heavy Higgs fields $H$, $\overline{H}$
which are assumed to gain \vevs{} \cite{King:OperatorAnalysis})
\begin{equation}
  \label{eq:heavy_higgs_\vevs}
  \left< H^{\alpha b} \right> = \left< \nu_H \right> = V
  \delta^\alpha_4 \delta^b_2 \sim M_X \; ;\; \left<
  \overline{H}_{\alpha x} \right> = \left< \overline{\nu}_H \right> =
  \overline{V} \delta_\alpha^4 \delta_x^2 \sim M_X
\end{equation}
This symmetry breaking splits the Higgs field $h$
into two Higgs doublets, $h_1$, $h_2$. Their neutral components then gain
weak-scale \vevs
\begin{equation}
  \label{eq:mssm_higgs_\vevs}
  \left< h_1^0 \right> = v_1 \; ; \;
  \left< h_2^0 \right> = v_2 \; ; \;
  \tan \beta = v_2 / v_1.
\end{equation}
The low energy limit of this model contains the MSSM with right-handed
neutrinos.  We will return to the right handed neutrinos when we
consider operators including the heavy Higgs fields $H$,
$\overline{H}$ which lead to effective Yukawa contributions and
effective Majorana mass matrices when the heavy Higgs fields gain
\vevs.

\subsection{Yukawa Operators}
\label{sec:yukawa-sector}

The Yukawa operators, responsible for generating effective Yukawa couplings,
have the following structure
\footnote{We note that due to the allocation
of charges, and since the effective Yukawa operators include the fields
$F_I \overline{F}_J h$ with overall
charge positive, the field
$\overline{\theta}$ does not enter the Yukawa operators.}
\cite{King:OperatorAnalysis}:
\begin{equation}
  \label{eq:dirac_operator_n_is_n}
  \mathcal{O} = F_I \overline{F}_J h \left(\frac{H
  \overline{H}}{M_X^2}\right)^n
  \left(\frac{\theta}{M_X}\right)^{p(i,j)}
\end{equation}
where the integer
$p(i,j)$ is the total $U(1)_F$ charge of $F_I + \overline{F}_J + h$ and
$H\overline{H}$ has a $U(1)_F$ charge of zero.
The tensor structure of the operators in
Eq.\ref{eq:dirac_operator_n_is_n} is
\begin{equation}
  \label{eq:dirac_mass_tensor}
  \left(\mathcal{O}\right)^{\alpha\rho y w}_{\beta\gamma x z} =
  F^{\alpha a} \overline{F}_{\beta x} h^y_a \overline{H}_{\gamma z}
  H^{\rho w} \theta^{p(i,j)}
\end{equation}
One constructs \cite{King:OperatorAnalysis}
$\SU{4}_{PS}$ invariant tensors $C^{\beta\gamma}_{\alpha\rho}$ that
combine $4$ and $\overline{4}$ representations of $\SU{4}_{PS}$ into
\boldmath $1$, $6$, $10$, $\overline{10}$ and $15$ \unboldmath
representations. Similarly we construct $\SU{2}_R$ tensors
$R^{xz}_{yw}$ that combine $\mathbf{2}$ representations of \SU{2} into
singlet and triplet representations. These tensors are contracted
together and into $\mathcal{O}^{\alpha\rho y w}_{\beta \gamma
x z}$ to create singlets of $\SU{4}_{PS}$, $\SU{2}_L$ and $\SU{2}_R$.
Depending on which operators are used, different
Clebsch-Gordan coefficients (CGCs) will emerge.

We look at two different models for the Yukawa sector, which we refer
to as model I and model II. 
The models represent different $\mathcal{O}(1)$ paramaters
$a,a^\prime,a^{\prime\prime}$ in the following operator texture
\cite{Blazek:2003wz}:
\begin{equation}
  \label{eq:operator_texture}
  \mathcal{O} 
  =
  \left[
    \begin{array}{ccc}
      ( a_{11} \mathcal{O}^{Fc} + a^{\prime\prime}_{11}
      \mathcal{O}''^{Ae} )\epsilon^5 & ( a_{12} \mathcal{O}^{Ee} +
      a^\prime_{12} \mathcal{O}'^{Cb} ) \epsilon^3 & ( a^\prime_{13}
      \mathcal{O}'^{Cf} + a^{\prime\prime}_{13}\mathcal{O}''^{Ee} )
      \epsilon \\ ( a_{21} \mathcal{O}^{Dc} ) \epsilon^4 & ( a_{22}
      \mathcal{O}^{Bc} + a_{22}^\prime \mathcal{O}'^{Ff} ) \epsilon^2
      & ( a_{23} \mathcal{O}^{Ee} + a^\prime_{23} \mathcal{O}'^{Bc} )
      \\ ( a_{31} \mathcal{O}^{Fc} ) \epsilon^4 & ( a_{32}
      \mathcal{O}^{Ac} + a_{23}^\prime \mathcal{O}'^{Fe} ) \epsilon^2
      & a_{33}
    \end{array}
  \right]
\end{equation}
where the operator nomenclature is defined in 
Appendix \ref{sec:n=1-operators}.
For convenience, from this point on, we define:
\begin{equation}
  \label{eq:20}
  \delta=\frac{H\overline{H}}{M_X^2}
\end{equation}
and
\begin{equation}
  \label{eq:10}
  \epsilon=\frac{\theta}{M_X}
\end{equation}
Eq.\ref{eq:operator_texture} then yields the effective Yukawa matrices
\begin{eqnarray}
  \label{eq:explicit_yu}
  Y^{u}(M_X) &=&
  \left[
    \begin{array}{ccc}
    a^{\prime\prime}_{11} \sqrt{2} \delta^3 \epsilon^5 & 
    a^\prime_{12} \sqrt{2} \delta^2\epsilon^3 &  
    a^\prime_{13}\frac{2}{\sqrt{5}} \delta^2\epsilon  \\
    0 & 
    a^\prime_{22} \frac{8}{5 \sqrt{5}}\delta^2\epsilon^2 
    & 0 \\
    0 
    & a^\prime_{32} \frac{8}{5} \delta^2 \epsilon^2 
    & a_{33}
    \end{array}
  \right]
  \\
  \label{eq:26}
  Y^{d}(M_X) &=&
  \left[
    \begin{array}{ccc}
      a_{11} \frac{8}{5}\delta \epsilon^5 & -a^\prime_{12} \sqrt{2}
      \delta^2\epsilon^3 & a^\prime_{13} \frac{4}{\sqrt{5}} \\ a_{21}
      \frac{2}{\sqrt{5}}\delta \epsilon^4 & (a_{22} \sqrt{\frac{2}{5}}
      \delta + a^\prime_{22} \frac{16}{5\sqrt{5}} \delta^2)\epsilon^2
      & a^\prime_{23} \sqrt{\frac{2}{5}} \delta^2 \\ a_{31}
      \frac{8}{5}\delta\epsilon^4 & a_{32} \sqrt{2} \delta \epsilon^2
      & a_{33}
    \end{array}
  \right]
  \\
  \label{eq:27}
  Y^{e}(M_X) &=&
  \left[
    \begin{array}{ccc}
    a_{11} \frac{6}{5}\delta \epsilon^5 & 0 & 0 \\ a_21
    \frac{4}{\sqrt{5}} \delta\epsilon^4 & ( -a_{22}
    3\sqrt{\frac{2}{5}} \sqrt{\frac{2}{5}} + a^\prime_{22} \delta
    \frac{12}{5\sqrt{5}} ) \delta\epsilon^2 & -a^\prime_{23}
    \sqrt{\frac{2}{5}} \delta^2 \\ -a_{31} \frac{6}{5}
    \delta\epsilon^4 & a_{23} \sqrt{2} \delta\epsilon^2 & a_{33}
  \end{array}
\right]
  \\
  \label{eq:28}
  Y^{\nu}(M_X) &=&
  \left[
    \begin{array}{ccc}
      a^{\prime\prime}_{22} \sqrt{2}\delta^3\epsilon^5 & 
      a_{12} 2 \delta\epsilon^3 & 
      a^{\prime\prime}_{13} \delta^3\epsilon \\
      0 & 
      a^\prime_{22} \frac{6}{5\sqrt{5}}\delta^2\epsilon^2 & 
      a_{23} 2 \delta \\
      0 & 
      a^\prime_{32} \frac{6}{5} \delta^2\epsilon^2 & 
      a_{33}
    \end{array}
  \right]
\end{eqnarray}

\begin{table}[tp]
  \centering
  \begin{tabular}{|c|c|c|}
    \hline
    & Model I & Model II \\
    \hline
    $a_{33}$ & 0.55 & 0.55 \\
    \hline
    $a_{11}$ & -0.92 & -0.92 \\
    $a_{12}$ & 0.33 & 0.33 \\
    $a_{21}$ & 1.67 & 1.67 \\
    $a_{22}$ & 1.12 & 1.12 \\
    $a_{23}$ & 0.89 & 0.89 \\
    $a_{31}$ & -0.21 & -0.21 \\
    $a_{32}$ & 2.08 & 2.08 \\
    \hline
    $a^\prime_{12}$ & 0.77 & 0.77 \\
    $a^\prime_{13}$ & 0.53 & 0.53 \\
    $a^\prime_{22}$ & 0.66 & 0.66 \\
    $a^\prime_{23}$ & 0.40 & 0.40 \\
    $a^\prime_{32}$ & 1.80 & 1.80 \\
    \hline
    $a^{\prime\prime}_{11}$ & 0.278 & 0.278 \\
    $a^{\prime\prime}_{13}$ & 0.000 & 1.000 \\
    \hline
    $A_{11}$ & 0.94 & 0.94 \\
    $A_{12}$ & 0.48 & 0.48 \\
    $A_{13}$ & 2.10 & 2.10 \\
    $A_{22}$ & 0.52 & 0.52 \\
    $A_{23}$ & 1.29 & 1.79 \\
    $A_{33}$ & 1.88 & 1.88 \\
    \hline
  \end{tabular}
  \caption{The $a,a^\prime$~and~$a^{\prime\prime}$ paramaters for
  model I and model II}
  \label{tab:a_ap_app_for_models}
\end{table}
The order unity coefficients $a_{ij},a'_{ij}$ of the operators are
adjusted to give a good fit to the quark and lepton masses and mixing
angles, and take the values given in
Table~\ref{tab:a_ap_app_for_models}.
Note that the two models differ only in the choice of $a^{\prime\prime}_{13}$,
which is taken to be zero in model I. 
Model I consequently has a lower rate for
$\mu\rightarrow e\gamma$, and model II has a higher
$\mu\rightarrow e\gamma$ rate due to the non-zero 13 element of the
neutrino Yukawa matrix, as can be understood from the
analytic results in \cite{Blazek:2002wq}.
The fits assume $\delta = \epsilon =0.22$.

\subsection{Majorana Operators}
\label{sec:majorana-fermions}

We are interested in Majorana fermions because they can contribute
neutrino masses of the correct order of magnitude via the see-saw
effect. The operators for Majorana fermions are of the form

\begin{equation}
  \label{eq:majorana_operators_n_is_n}
  \mathcal{O}_{IJ} = \overline{F}_I \overline{F}_J \left(\frac{H H}{M_X}\right)
  \left(\frac{H \overline{H}}{M_X^2}\right)^{n-1}
  \left(\frac{\theta}{M_X}\right)^{q_{IJ}}
\end{equation}

There do not exist renormalisable elements of this infinite series of
operators, so $n < 1$ Majorana operators are not defined
\footnote{Except for the 33 neutrino mass term; this is allowed
because of string theoretic effects}. A similar analysis goes through
as for the Dirac fermions; however the structures only ever give
masses to the neutrinos, not to the electrons or to the
quarks. \footnote{ To see this note that the form of the two H \vevs{}
is symmetric, and proportional to $\delta^\alpha_4
\delta^x_2$. Symmetric structures will then contract to give neutrino
mass terms. Antisymmetric structures will contract to give zero. As
any structure can be written as a sum of a symmetric and an
antisymmetric part, we see immediately that the only mass terms can be
given to the neutrinos because of the form of the \vevs{} in
Eq.~(\ref{eq:heavy_higgs_vevs}) }

It should be noted that the Majorana neutrinos will not affect the
A-terms, as these operators do not contribute to the Yukawas.
The RH Majorana neutrino mass matrix is:
\begin{equation}
  \label{eq:21}
  \frac{M_{RR}(M_X)}{M_{33}} =
  \left[
    \begin{array}{ccc}
      A_{11}\delta\epsilon^8 & A_{12}\delta\epsilon^6 &
      A_{13}\delta\epsilon^4 \\ A_{12}\delta\epsilon^6 &
      A_{22}\delta\epsilon^4 & A_{23}\delta\epsilon^2 \\
      A_{13}\delta\epsilon^4 & A_{23}\delta\epsilon^2 & A_{33}
    \end{array}
  \right]
\end{equation}

\subsection{Neutrino sector results}
\label{sec:neutr-sect-results}

The neutrino Yukawa matrix in Eq.\ref{eq:28} and the heavy Majorana 
mass matrix in Eq.\ref{eq:21} imply that the see-saw mechanism 
satisfies the condition of sequential dominance \cite{SRHND},
leading to a natural neutrino mass hierarchy $m_1\ll m_2\ll m_3$
with no fine-tuning.
The dominant contribution to the atmospheric neutrino mass 
$m_3$ comes from the third (heaviest) right-handed neutrino,
with the leading subdominant contribution to the solar
neutrino mass $m_2$ 
coming from the second right-handed neutrino.
In such a natural scenario, the large atmospheric angle
is due to the large ratio of dominant neutrino Yukawa couplings
$\tan \theta_{23}\approx Y^{\nu}_{23}/Y^{\nu}_{33}$,
and the large solar angle is due to the large ratio
of leading subdominant Yukawa couplings
$\tan \theta_{12}\approx \sqrt{2} Y^{\nu}_{12}/(Y^{\nu}_{22}-Y^{\nu}_{32})$.

Model I for the Yukawa sector is taken from a global
analysis of a SUSY Pati-Salam model enhanced with an Abelian flavour
symmetry \cite{Blazek:2003wz}. At one-loop
order the Yukawa runings only depend on the other Yukawas and the
gauge couplings. Since Model II only differs from Model I in the
neutrino Yukawa, we do not expect the quark masses or mixing angles to
be different. We also do not expect the charged lepton masses to differ by
much.

\begin{table}[htbp]
  \centering
  \begin{tabular}{|c||c|c|c|}
    \hline Observable & Model I & Model II & Experimental \\ &
    Prediction & Prediction & Values \\ \hline $\sin^2\theta_{12}$ &
    $0.316$ & $0.308$ & $0.28 \pm 0.05$ \\ $\sin^2\theta_{23}$ &
    $0.553$ & $0.552$ & $0.50 \pm 0.15$ \\ $\sin^2\theta_{13}$ &
    $5.18\cdot10^{-3}$ & $5.20\cdot10^{-3}$ & $< 0.03$ \\ \hline
    $\Delta m^2_{atm}$ & $1.32\cdot 10^{-3}$ & $1.33\cdot10^{-3}$ &
    $(2.5\pm0.8)10^{-3}$ \\ $\Delta m^2_{sol}$ & $6.05\cdot 10^{-5}$ &
    $5.91\cdot10^{-5}$ & $(7.0\pm3.0)10^{-5}$ \\ \hline
  \end{tabular}
  \caption{The neutrino mass differences and mixing angles in model I,
  model II and the experimental limits}
  \label{tab:neutrino_data}
\end{table}

The possibility remains open, however, that the new operator in the 13
Yukawa elements could predict either a mass difference or a neutrino
mixing angle in violation of the results from the various neturino
experiments \cite{GoPe, SKamiokandeColl, SNO:results, CHOOZ:results}. As such, we
checked our predictions for the mass-differences and the mixing angles
for both models, in comparison to experiment. The results of this are
summarised in Table~\ref{tab:neutrino_data}.

We note that in both model I and model II, we are within the
constraints on the neutrino sector. In fact 
Model II is slightly closer to the central values of
three observable paramaters (the solar and atmospehric neutrino
mixing angles, and the atmospheric mass difference). In both cases
we predict values of $\theta_{13}$ below the current limit.

\section{Soft Supersymmetry Breaking Masses}
\label{soft}

\subsection{Supersymmetry Breaking F-terms}

In \cite{Everett:2002pm} it was assumed that the Yukawas were
field-independent, and hence the only $F$-vevs of importance were that
of the dilaton ($S$), and the untwisted moduli ($T^i$).
Here we set out the paramaterisation for the F-term \vevs{},
including the contributions from the FN field $\theta$ and the
heavy Higgs fields $H,\overline{H}$. Note that
the field dependent part follows from the assumption that the
family symmetry field, $\theta$ is an intersection state.
\begin{eqnarray}
  \label{eq:dilaton_auxilliary_vev_42241}
  F_S
  &=&
  \sqrt{3} m_{3/2} \left( S + \overline{S} \right)
  X_S
  \\
  \label{eq:untwisted_mod_auxilliary_vev_42241}
  F_{T_i}
  &=&
  \sqrt{3} m_{3/2} \left( T_i + \overline{T}_i \right)
  X_{T_i}
  \\
  \label{eq:heavy_higgs_auxilliary_vev_42241}
  F_{H^{\alpha b}} 
  &=&
  \sqrt{3} m_{3/2} H^{\alpha b} \left( S + \overline{S} \right)^\frac{1}{2}
  X_H
  \\
  \label{eq:heavy_conj_higgs_auxilliary_vev_42241}
  F_{\overline{H}_{\alpha x}} &=& \sqrt{3} m_{3/2}
  \overline{H}_{\alpha x} \left( T_3 + \overline{T}_3
  \right)^\frac{1}{2} X_{\overline{H}} \\
  \label{eq:family_field_auxilliary_vev_42241}
  F_\theta
  &=&
  \sqrt{3} m_{3/2} \theta \left(S + \overline{S}\right)^\frac{1}{4}
  \left(T_3 + \overline{T}_3\right)^\frac{1}{4}
  X_\theta
\end{eqnarray}
We introduce a shorthand notation:
\begin{equation}
  \label{eq:shorthand_key}
  F_H H = \sum_{\alpha b} F_{H^{\alpha b} } H^{\alpha b} \; ;\;
  F_{\overline{H}} \overline{H} = \sum_{\alpha x}
  F_{\overline{H}_{\alpha x} } \overline{H}_{\alpha x}.
\end{equation}

\subsection{Soft Scalar Masses}
\label{sec:scalars_42241}
There are two contributions to scalar mass
squared matrices, coming from SUGRA and from D-terms.
In this subsection we calculate the SUGRA predictions for the
matricies at the GUT scale, and in the next subsection we add on the 
D-term contributions.

The SUGRA contributions to soft masses are detailed in 
Appendix~\ref{sec:soft-from-SUGRA}.
From Eq.~(\ref{eq:normalised_scalars}) we
can get the family independent form for all scalars:
\begin{eqnarray}
  \label{eq:scalars_left}
  m^2_{L} =
  &m^2_{3/2}&
  \left[
    \begin{array}{ccc}
      a & & \\
      & a & \\
      & & b_L
    \end{array}
  \right] \\
  m^2_{R} = 
  \label{eq:scalars_right}
  &m^2_{3/2}&
  \left[
    \begin{array}{ccc}
      a & & \\
      & a & \\
      & & b_R
    \end{array}
  \right] \\
  \label{eq:mssm_higgs_at_mx}
  m^2_h =  & m^2_{3/2}& ( 1 - 3 X^2_S )\\
  \label{eq:ps_H_at_mx}
  m^2_H = & m^2_{3/2}& ( 1 - 3 X^2_S ) \\
  \label{eq:ps_Hbar_at_mx}
  m^2_{\overline{H}} =& m^2_{3/2}&(1 - 3 X^2_{T_3} )
\end{eqnarray}
where
\begin{eqnarray}
  a &=& 1 - \frac{3}{2} (X^2_S + X^2_{T_3} ) \\
  b_L &=& 1 - 3 X^2_{T_3} \\
  b_R &=& 1 - 3 X^2_{T_2}
\end{eqnarray}
Here $m^2_L$ represents the left handed scalar mass squared
matricies $m^2_{QL}$ and $m^2_{LL}$. $m^2_R$ represents the right
handed scalar mass squared matricies $m^2_{UR}$, $m^2_{DR}$,
$m^2_{ER}$ and $m^2_{NR}$.

\subsection{D-term Contributions}

We now consider the
D-terms from breaking the Pati-Salam group \psgroup down to the
MSSM group \smgroup. These will be family independent, but charge
dependent, and will pull the six matricies that appear in the RGE
equations apart. We shall neglect D-term contributions from the broken
family symmetry which would lead to additional sources of flavour violation. 

The addition to the D-terms have been written down before
\cite{King:2000vp}. The corrections are, in matrix notation:
\begin{eqnarray}
  \label{eq:11}
  m^2_{QL} &=& m^2_{L} + g_4^2 D^2 \\
  \label{eq:12}
  m^2_{UR} &=& m^2_{R } - ( g_4^2 - 2 g^2_{2R} ) D^2 \\
  \label{eq:13}
  m^2_{DR} &=& m^2_{R } - ( g_4^2 + 2 g^2_{2R} ) D^2 \\
  \label{eq:14}
  m^2_{LL} &=& m^2_{L} - 3 g_4^2 D^2 \\
  \label{eq:15}
  m^2_{ER} &=& m^2_{R} + ( 3 g_4^2 - 2g^2_{2R} )D^2 \\
  \label{eq:16}
  m^2_{NR} &=& m^2_{R} + ( 3 g_4^2 + 2g^2_{2R} )D^2 \\
  \label{eq:17}
  m^2_{h_u} &=& m^2_{h_2} - 2g_{2R}^2 D^2\\
  \label{eq:18}
  m^2_{h_d} &=& m^2_{h_1} + 2g_{2R}^2 D^2
\end{eqnarray}
where in the appendix of Ref.\cite{King:2000vp}, an expression for $D^2$ in
terms of the soft paramaters $m^2_H$ and $m^2_{\overline{H}}$ is
derived,
\begin{eqnarray}
  \label{eq:19}
  D^2 = \frac{m^2_{H} - m^2_{\overline{H}} }{4 \lambda_S^2 + 2g_{2R}^2
  + 3g_4^2 }.
\end{eqnarray}
The gauge couplings and mass parameters in Eq.\ref{eq:19}
are predicted from the model. The only free parameter is
the coupling $\lambda_S$ is a dimensionless coupling constant which
enters the potential \cite{King:2000vp} and 
should be perturbative. Furthermore, we see that the largest
that $D^2$ can be is when $\lambda_S$ is zero, so not only is the order of
magnitude of $D^2$ predicted in this model, but we also have an exact upper
bound on the value.

\subsection{Soft Gaugino Masses}
\label{sec:param-gaug}
The soft gaugino masses are the same as in
\cite{Everett:2002pm}, which we quote here for completeness.
The results follow from
Eq.~(\ref{eq:normalised_gauginos}) applied to the \psgroup gauginos, which
then mix into the \smgroup gauginos whose masses are given by
\begin{eqnarray}
  \label{eq:su3_gaugino_gut_scale}
  M_3 &=& \frac{\sqrt{3} m_{3/2} } {\left(T_1 + \overline{T}_1\right)
  + \left(T_2 + \overline{T}_2\right) } \left[ (T_1 + \overline{T}_1)
  X_{T_1} + (T_2 + \overline{T}_2 ) X_{T_2} \right] \\
 \label{eq:su2_gaugino_gut_scale}
  M_2 &=& \sqrt{3} m_{3/2} X_{T_1} \\
  \label{eq:u1_gaugino_gut_scale}
  M_1 &=& \frac{\sqrt{3} m_{3/2} } {\frac{5}{3} (T_1 + \overline{T}_1
  ) + \frac{2}{3} (T_2 + \overline{T}_2 ) } \left[ \frac{5}{3}(T_1 +
  \overline{T}_1) X_{T_1} + \frac{2}{3}(T_2 + \overline{T}_2 ) X_{T_2}
  \right]
\end{eqnarray}
The values of $T_1 + \overline{T}_1$ and
$T_2 + \overline{T}_2$ are proportional to the brane
gauge couplings $g_{5_1}$ and $g_{5_2}$, which are related in a simple
way to the MSSM couplings at the unification scale. This is discussed
in \cite{Everett:2002pm}.

When we run the MSSM gauge couplings up and solve for $g_{5_1}$ and
$g_{5_2}$ we find that approximate gauge coupling unification is 
achieved by $T_1 +\overline{T_1} \gg T_2 +
\overline{T}_2$. Then we find the simple approximate result
\begin{equation}
  \label{eq:63}
  M_1 \approx M_3 \approx M_2 = \sqrt{3}m_{3/2} X_{T_1}.
\end{equation}

\subsection{Soft Trilinear Masses}
\label{sec:param_tril}
So far the soft masses are as in \cite{Everett:2002pm},
with the FN fields and heavy Higgs contributions being completely
negligible due to the smallness of their F-terms.
However for the soft trilinear masses 
these contributions are of order $O(m_{3/2})$ despite 
having small F-terms, so FN and Higgs contributions will
give very important additional contributions beyond those
considered in \cite{Everett:2002pm}.

From Appendix~\ref{sec:soft-from-SUGRA} we see that
the canonically normalised equation for the trilinear is:
\begin{equation}
  \label{eq:normalised_trilinear0}
  A_{abc} = F_I 
  \left[
    \overline{K}_I - \partial_I \ln
    \left(
      \tilde K_a \tilde K_b \tilde K_c
    \right)
  \right]
  + F_m \partial_m \ln Y_{abc}
\end{equation}
This general form for the
trilinear accounts for contributions from non-moduli F-terms. These
contributions are in general expected to be of the same magnitude as
the moduli contributions despite the fact that the non-moduli F-terms
are much smaller \cite{Abel:2001cv}. Specifically,
if the Yukawa hierarchy is taken to be generated by a FN
field, $\theta$ such that $Y_{ij} \sim \theta^{p_{ij}}$, 
then we expect $F_\theta \sim m_{3/2} \theta$, 
and then $\Delta A_{ij}=F_\theta \partial_\theta \ln Y_{ij} \sim
p_{ij}m_{3/2}$ and so even though these fields are expected to have heavily
sub-dominant F-terms
\footnote{In our model the FN and heavy Higgs vevs are 
of order the unification scale, compared to the moduli vevs
which are of order the Planck scale.}
they contribute to the trilinears at the same
order $O(m_{3/2})$ as the moduli, but in a flavour off-diagonal way.

In the specific D-brane model of interest here the general results for
soft trilinear masses, including the contributions for general effective Yukawa
couplings are given in Appendix \ref{sec:param-tril-42241}.
From Eqs.\ref{eq:dirac_operator_n_is_n},\ref{eq:dirac_mass_tensor}
we can read off the effective Yukawa couplings,
\begin{equation}
  \label{eq:dirac_yukawa}
  Y_{h F \overline{F} } h F \overline{F} \equiv
  \underbrace{(c)^{\beta\gamma}_{\alpha\rho}(r)^{xz}_{yw}\overline{H}_{\gamma
  z} H^{\rho w}\theta^{p}}_{ {Y_{hF\overline{F}}}^{\beta
  x}_{\alpha y}}h^y_a F^{\alpha a} \overline{F}_{\beta x}.
\end{equation}
Note the extra group indicies that the effective Yukawa coupling
${Y_{hF\overline{F}}}^{\beta x}_{\alpha y}$ has, and 
proper care must be taken of the tensor structure
when deriving trilinears from a given operator.
For Model I and II defined earlier, we can write down the
trilinear soft masses, $A$, by substituting the operators in
Eq.\ref{eq:operator_texture} into the results in 
Appendix \ref{sec:param-tril-42241}.
Having done this we find the result:
\begin{equation}
  \label{eq:trilinear}
  A = \sqrt{3} m_{3/2}
  \left[
    \begin{array}{ccc}
      d_1 + d_H + 5d_\theta & d_1 + d_H + 3d_\theta & d_2 + d_H + d_\theta \\
      d_1 + d_H + 4d_\theta & d_1 + d_H + 2d_\theta & d_2 + d_H \\
      d_3 + d_H + 4d_\theta & d_3 + d_H + 2d_\theta & d_4
    \end{array}
  \right]
\end{equation}
where
\begin{eqnarray}
    d_1 &=& X_S - X_{T_1} - X_{T_2} \\ d_2 &=& \frac{1}{2} X_S -
  X_{T_1} - \frac{1}{2}X_{T_2} \\ d_3 &=& \frac{1}{2} X_S - X_{T_1} -
  X_{T_2} + \frac{1}{2}X_{T_£} \\ d_4 &=& -X_{T_1} \\ d_H &=&
  (S+\overline{S})^\frac{1}{2} X_H + (T_3 +
  \overline{T}_3)^\frac{1}{2} X_{\overline{H}} \\ d_\theta &=&
  (S+\overline{S})^\frac{1}{4} (T_3 + \overline{T}_3)^\frac{1}{4}
  X_\theta
\end{eqnarray}

\subsection{Why we expect large flavour violation}
\label{sec:high_fv}

According to ref.\cite{Kobayashi:2000br}, if the trilinears can
be written in a certain manner, then flavour violation is expected to
be small. However, it is not possible to write the trilinears in this
manner if either $X_H \ne 0$ or $X_{\overline{H}} \ne 0$ or $X_\theta
\ne 0$. It is possible in general to write:
\begin{equation}
  \label{eq:85}
  A_{ij} = A^L_i + A^R_j + \delta_{ij}
\end{equation}
If $\delta_{ij} = 0$, then the trilinears factorise. 
\begin{equation}
  \label{eq:86}
  \tilde{A} = 
  \left[
    \begin{array}{ccc}
      & & \\
      & Y_{ij} & \\
      & &
    \end{array}
  \right]
  \left[
    \begin{array}{ccc}
      A^R_1 & & \\
       & A^R_2 & \\
       & & A^R_3
    \end{array}
  \right]
  +
  \left[
    \begin{array}{ccc}
    A^L_1 & & \\
    & A^L_2 & \\
    & & A^L_3
  \end{array}
\right]
  \left[
    \begin{array}{ccc}
    &&\\
    & Y_{ij} & \\
    & &
  \end{array}
\right]
\end{equation}
If this is true at the SUSY breaking scale, then the FCNC effects are
small, and the leading order contributions are proportional to $A^R_i
- A^R_j$ or $A^L_i - A^L_j$.
If there are any contributions which are universal, $A^0$, then we can
add them in any linear combination to $A^L_i$ and $A^R_j$ provided that
$A^{\prime L}_i + A^{\prime R}_j = A^L_i + A^R_j + A^0$.
From the SUGRA formula for $A_{ij}$, it is clear which terms contribute
to the universal part, the left part and the right part:
\begin{equation}
  \label{eq:87}
  A_{ij} = A_{hij} = \underbrace{F^a\partial_a (\tilde{K} - \ln K^h_h )}_{A^0}
  - \underbrace{F^a\partial_a \ln K^i_i}_{A^L_i} 
  - \underbrace{F^a\partial_a \ln K^j_j}_{A^R_j} 
  + \delta_{ij}
\end{equation}
We see that $\delta_{ij}$ is the terms due to the derivative of the
Yukawa. If this is either zero or universal, then the $A$ matrix can
be written in the restricted form.

Unfortunately, neither the Higgs contribution or the Froggatt-Nielsen
contribution can be written in this form. The Higgs contribution 
to Eq.\ref{eq:trilinear} is:
\begin{equation}
  \label{eq:88}
  A_{H} \propto 
  \left[
    \begin{array}{ccc}
      a & a & a \\
      a & a & a \\
      a & a & b
    \end{array}
  \right]
  = a + (b-a)
  \left[
    \begin{array}{ccc}
      0 & 0 & 0 \\
      0 & 0 & 0 \\
      0 & 0 & 1
    \end{array}
  \right]
\end{equation}
The Froggatt-Nielsen contribution is maximally non-universal, and the
elements have $A_{ij} \;/\!\!\!\!\!\!\approx A_{kl}$ for $i \ne k\; ;
\; j \ne l$. In this case we expect there to be the largest
contribution to flavour violation, assuming that it is not tuned
down. ( We could do this either by selecting a very small value for
the F-term \vev{} by setting $X_\theta \approx 0$, or by setting the
operator texture in the Yukawas to have very small off-diagonal
elements, as the A-contribution multiplies Yukawa elements ).
Hence we see that the new sources of flavour violation will not only
contribute to the trilinear terms on at least an equal footing as the
moduli, but that also they cannot be written in a form where the
contribtuion to flavour violation is expected to be small.

\section{Results}
\label{sec:numerical-etc} 

\subsection{Benchmark points}

Since the parameter space for this model is reasonably expansive, and
the intention is to compare different sources of LFV, it is convenient
to consider four benchmark points, as follows.
It should be noted that for all these points, we have taken all $X_{T_i}$
to be the same, $X_{T_i} = X_T$, and also $X_H = X_{\overline{H}}$.

\label{sec:four_points}
\begin{table}[htbp]
  \centering
  \begin{tabular}{|c|c|c|c|c|c|}
    \hline
    Point & $X_S$ & $X_T$ & $X_H$ & $X_{\overline{H}}$ & $X_\theta$ \\
    \hline
    A & 0.500 & 0.500 & 0.000 & 0.000 & 0.000 \\
    B & 0.536 & 0.488 & 0.000 & 0.000 & 0.000 \\
    C & 0.270 & 0.270 & 0.000 & 0.000 & 0.841 \\
    D & 0.270 & 0.270 & 0.578 & 0.578 & 0.000 \\
    \hline
  \end{tabular}
  \caption{The four benchmark points, A-D}
  \label{tab:benchmarks_defined}
\end{table}
\begin{itemize}
\item 
  Point A is referred to as ``minimum flavour violation''.  At the
  point $X_S=X_{T_i}$ the scalar mass matrices $m^2$ are proportional
  to the identity, and the trilinears $\tilde A$ are aligned with the
  Yukawas. Also, if we look back to eq.~(\ref{eq:19}),
  eq.~(\ref{eq:ps_H_at_mx}) and eq.~(\ref{eq:ps_Hbar_at_mx}) we see
  that for $X_S = X_T$, which is the case for point A ( and point C,
  and point D ) we see that the upper limit on the magnitude of the
  D-term contribution is zero.  As such both $m^2$ and $\tilde{A}$
  will be diagonal in the SCKM basis in the abscence of the RH
  neutrino field.
  
\item 
  Point B is referred to as ``SUGRA''. With $X_S\neq X_{T_i}$ it
  represents typical flavour violation from the moduli fields; this is
  the amount of flavour violation that would traditionally have been
  expected with no contribution from the $F_H$ or $F_\theta$ fields.
  
\item 
  Point C is referred to as ``FN flavour violation''.  It represents
  flavour violation from the Froggatt-Nielsen sector by itself,
  without any contribution to flavour violation from traditional SUGRA
  effects since $X_S=X_{T_i}$ as in point A.
  
\item 
  Point D is referred to as ``Heavy Higgs flavour violation''.  It
  represents flavour violation from the heavy Higgs sector, without
  any contribution from either traditional SUGRA effects since
  $X_S=X_{T_i}$, or from FN fields sinc $F_\theta=0$.  As will become
  apparent, at this point the seesaw mechanism actually helps reduce
  the LFV for $\mu\rightarrow e\gamma$ in model I and
  $\tau\rightarrow\mu\gamma$ in both models.
\end{itemize}
\subsection{Numerical Results}

From the benchmark points defined in Table~\ref{tab:benchmarks_defined}, 
the F-term \vevs{} were determined, and
from these the soft paramaters at the high energy scale $M_X =
3.10^{16}\mathrm{GeV}$ were calculated. The soft paramaters were then
run down using the 1-loop RGEs of the MSSM + $\nu^c$ model. 
For our numerical results we 
use a modified version of SOFTSUSY \cite{Allanach:2001kg}. The
modifications were made to add the effect of the right-handed neutrino
field to the RGEs and to decouple them in a manner that allows the
neutrino masses and mixing angles to be calculated at the low energy
scale. As a result of the RGEs having to be recoded, all of them are
to one loop only in the version that was used here.

Flavour violation is proportional to non-zero off diagonal elements in
the scalar mass squared matricies $m^2$ in the SCKM basis and to
non-zero off diagonal elements in the trilinears $\tilde A$ in the
SCKM basis. 
\footnote{The SCKM basis is the basis where the yukawas are
diagonal at the electroweak scale}. Hence, there are two ways to
generate flavour violation. The first is to have non-zero off-diagonal
elements in $m^2$ of the scalars and $\tilde{A}$ at the unification
scale. The second is to have non-zero off diagonal elements
radiatively generated by the $\beta$-function running down to the
electroweak scale. It is possible to remove the second source by
removing the RH neutrino field from the model; this allows a
disentangling of the see-saw mechanism from the particular source
of interest, but is unphysical since we know that
the neutrinos have to be massive.

\begin{figure}[tp]
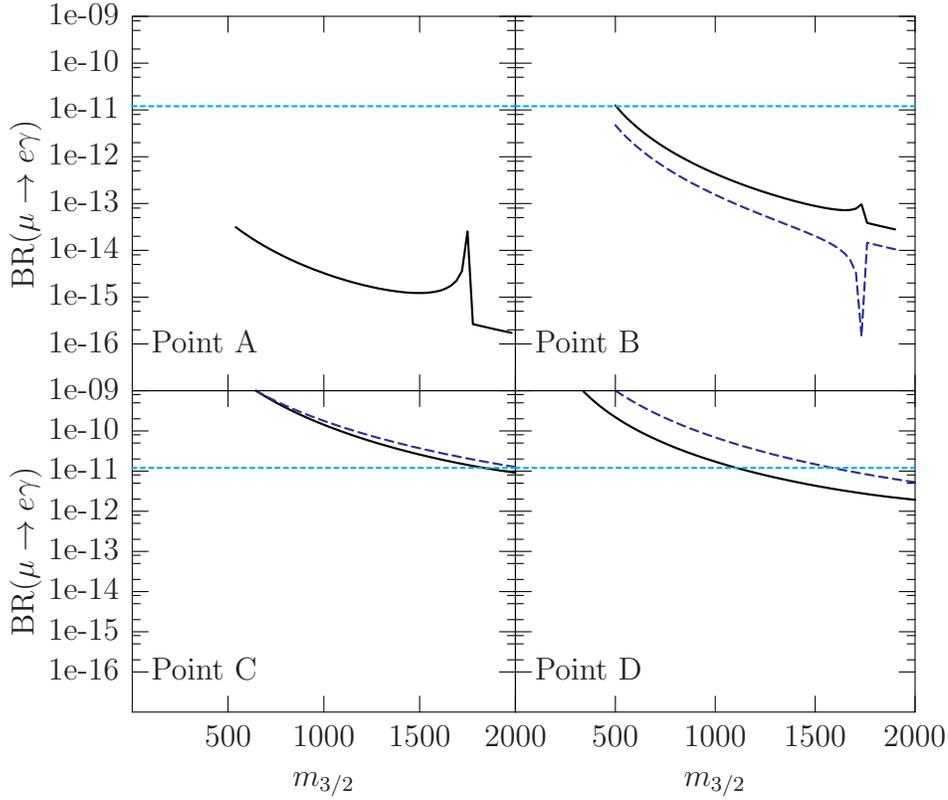
 \include{m32_meg_lmeg}\centering  
  \caption{$\mathrm{BR}(\mu\rightarrow e\gamma)$ for points A-D in
  model I ( low $\mu\rightarrow e\gamma)$. The solid line represents
  model I. The dashed line represents an unphysical model with no
  right-handed neutrino field whose purpose is only comparison. The
  horizontal line is the 2002 experimental limit from
  ref.\cite{Hagiwara:fs}}.
  \label{fig:meg_model_I}
\end{figure} 

\begin{figure}[bp]
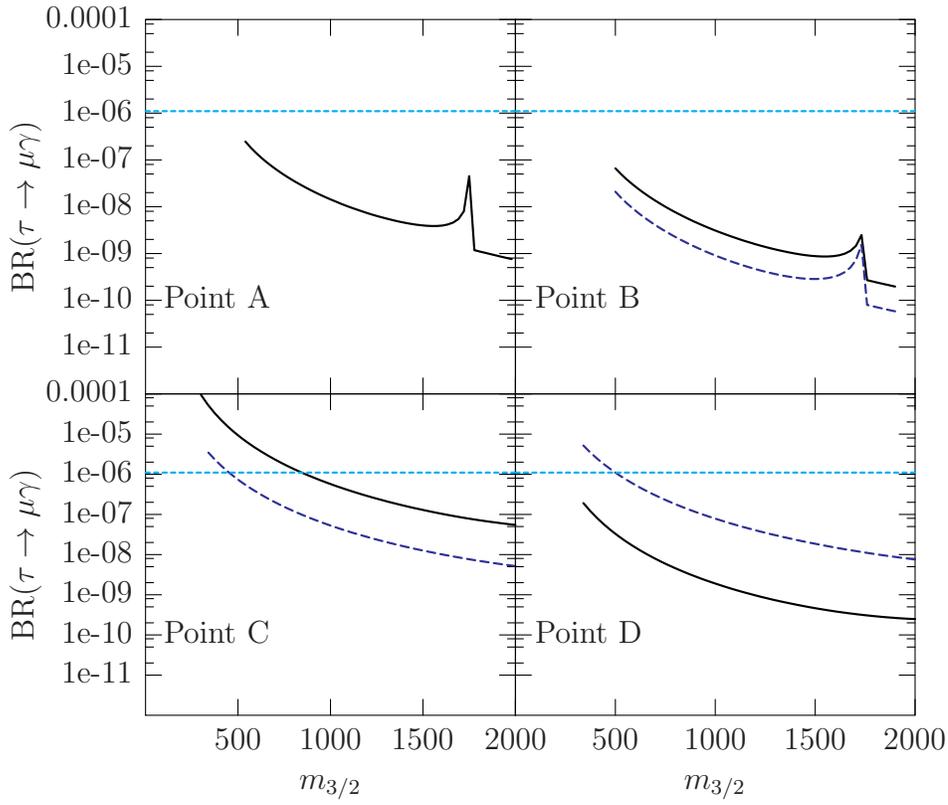

\include{m32_tmg} 
\centering
\caption{$\mathrm{BR}(\tau\rightarrow\mu\gamma)$ for points A-D. The
lines coincide in both model I and model II.  The solid line
represents models I and II ( which predict very similar rates for
$\tau\rightarrow\mu\gamma$).  The dashed line represents an unphysical
model with no right-handed neutrino field whose purpose is only
comparison. The horizontal line is the 2002 experimental limit from
ref.\cite{Hagiwara:fs}}
\label{fig:tmg_both_models}
\end{figure}

\begin{figure}[tp]
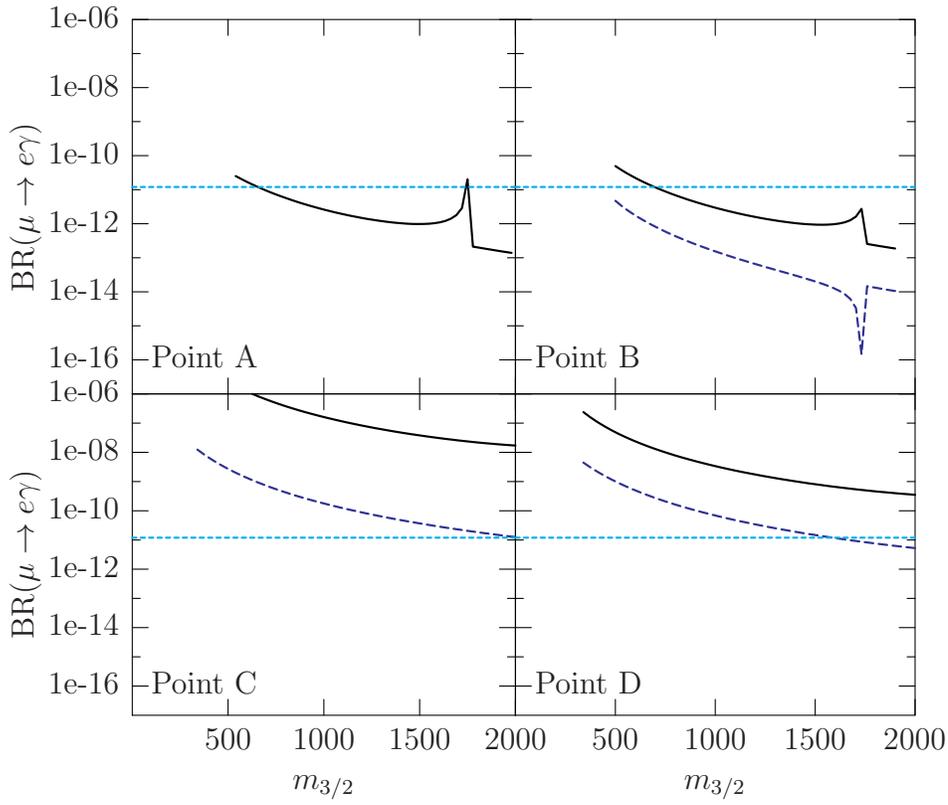

\include{m32_meg_hmeg}
\centering
\caption{$\mathrm{BR}(\mu\rightarrow e\gamma)$ for points A-D in model
II (high $\mu\rightarrow e\gamma)$. Note that because the rate is much
higher the scale is different to that in
fig.~\ref{fig:meg_model_I}. The solid line represents model II. The
dashed line represents an unphysical model with no right-handed
neutrino field whose purpose is only comparison. The horizontal line
is the 2002 experimental limit from ref.\cite{Hagiwara:fs} }
\label{fig:meg_model_II}
\end{figure}

Figure \ref{fig:meg_model_I} shows numerical results for 
$\mathrm{BR}(\mu\rightarrow e\gamma)$ for Model I, 
plotted against the gravitino mass $m_{3/2}$,
where each of the four panels corresponds to each of the
four benchmark points A-D. As $m_{3/2}$
increases the sparticle spectrum becomes heavier. This will look different
at each paramater point, but the physical masses are expected to be of
the same order of magnitude as the gravitino mass. As such, high
gravitino masses will start to reintroduce the fine-tuning problem of
the gluino mass being too high.
Point A corresponds to minimum flavour violation, where the
only source of LFV is from the see-saw mechanism, which for 
Model I is well below the experimental limit, shown as the faint 
horizontal dashed line.
Point B has LFV arising from SUGRA, with the FN and heavy Higgs
sources of LFV switched off, and in this case we also show the results
with the see-saw mechanism switched off (dashed curve)
as well as with the see-saw mechanism with 
SUGRA contributions to LFV (solid curve).
In both cases the results are below the experimental limit for
$m_{3/2}$ above 500 GeV.
Point C is the FN benchmark point, and for this case we see that
the experimental limit is violated over the entire range
of $m_{3/2}$ shown, with the see-saw mechanism making very 
little difference. Point D shows the heavy Higgs point, for 
which the experimental limit 
is violated for $m_{3/2}$ below 1000 GeV. Interestingly, the 
effect of switching off the see-saw mechanism in this case 
(dashed curve) is to increase the rate 
for $\mathrm{BR}(\mu\rightarrow e\gamma)$.

Figure \ref{fig:tmg_both_models} shows results for
$\mathrm{BR}(\tau\rightarrow\mu\gamma)$ for Model I, 
plotted against the gravitino mass $m_{3/2}$.
Point A for minimum flavour violation is below the
experimental limit, as is point B corresponding to 
SUGRA, with the see-saw mechanism switched off corresponding as
before to the dashed curve.
Point C corresponding to FN violates the experimental limit
for lower $m_{3/2}$, with a rather large effect coming from the
see-saw mechanism. Point D shows the heavy Higgs point,
with the effect of the see-saw mechanism being to reduce
$\mathrm{BR}(\tau\rightarrow\mu\gamma)$ in conjunction with the
LFV coming from heavy Higgs, similar to the analagous effect observed
previously.

Figure \ref{fig:meg_model_II} shows the analagous results for 
$\mathrm{BR}(\mu\rightarrow e\gamma)$ for Model II.
As expected model II, which is supposed to give a
high rate for $\mu\rightarrow e\gamma$, does give results close to the
experimental limit for points $A$ and $B$, and the limit is now
well exceeded for points $C$ and $D$. 
By increasing the gravitino mass sufficiently (which increases 
all the sparticle masses) it is possible to respect
the current experimental limit, but at the expense of a very heavy
superpartner spectrum. However it is worth noting that for 
benchmark points C,D the value of $X_T$ is almost half its value
corresponding to points A,B. According to Eq.\ref{eq:63} this
implies that for points C,D the gaugino masses are almost half their
values corresponding to points A,B, leading to reduced
fine-tuning for a given $m_{3/2}$.

The results for $\mathrm{BR}(\tau\rightarrow\mu\gamma)$ for Model II
are almost identical to those shown for Model I in
Figure \ref{fig:tmg_both_models}, which is as expected since the
only difference between the two models is in the 13 element of 
the neutrino Yukawa matrix.

\section{Conclusions} 
\label{sec:conclusions} 

We have investigated the phenomenological impact of different sources
of lepton flavour violation arising from realistic D-brane inspired
models based on supergravity mediated supersymmetry breaking,
where the origin of flavour is due to Froggatt-Nielsen (FN) operators.
We have discussed four distinct sources of lepton flavour violation
in such models: minimum flavour violation, arising from neutrino
masses and the see-saw mechanism with renormalisation group (RG)
running; supergravity flavour violation due to the non-universal
structure of the supergravity model; FN flavour violation due to 
the FN fields developing supersymmetry breaking F-terms and
contributing in an non-universal way to soft trilinear terms;
and finally heavy Higgs flavour violation arising from the 
Higgs fields used to break the unified gauge symmetry which may
behave analagously to the FN fields. 

In order to quantify the relative
effects, we studied a particular type I string inspired model based
on a supersymmetric Pati-Salam model arising from intersecting
D-branes as proposed in \cite{Everett:2002pm}, but here 
supplemented by a $U(1)$ family symmetry with the quarks and 
leptons described by the set of FN operators as in \cite{Blazek:2003wz}.
We have derived the soft supersymmetry breaking masses for the
model, including the new flavour vilating contributions to the soft 
trilinear masses arising from the FN and heavy Higgs fields.
We then performed a numerical analysis of LFV for four benchmark points, 
each chosen to highlight a particular source of flavour violation,
with the benchmark points $C$ and $D$ coresponding to LFV arising
from the FN and heavy Higgs fields giving by far the largest effects.
Since the new contributions are dominantly from the trilinears $\tilde A$, 
the amount of flavour violation is therefore
strongly dependent on the choice of Yukawa matrices at the
unification scale. For example
the huge difference in the rate of $\mu\rightarrow
e\gamma$ between Model I and Model II is simply
generated by changing the (1,3) element of $Y^\nu$.
Also we find that $\mu\rightarrow e\gamma$ is more constraining 
than $\tau\rightarrow\mu\gamma$. 

The numerical results show that the contributions to LFV
from Yukawa operators with the heavy Higgs sector and the
Froggatt-Nielsen sector can give the dominant contributions to LFV
processes, greatly exceeding contributions from SUGRA and
the see-saw mechanism, and should be taken into account when performing
phenomenological analyses of supergravity models.

\vskip 0.1in
\noindent
{\large {\bf Acknowledgements}}\\
S.K. thanks PPARC for a Senior Fellowship and I.P. thanks PPARC for a studentship.
We are also grateful to J. Parry and T. Bla\v{z}ek for discussions and also
for contributing some code. 
\newpage

\appendix

\section{Soft terms from supergravity}
\label{sec:soft-from-SUGRA}

We summarise here the standard way of getting soft SUSY breaking terms
from supergravity.  Supergravity is defined in terms of a K\"ahler
function, $G$, of chiral superfields ($\phi = h, C_a$). Taking the
view that the supergravity is the low energy effective field theory
limit of a string theory, the hidden sector fields $h$ are taken to
correspond to closed string moduli states ($h = S, T_i$), and the
matter states $C_a$ are taken to correspond to open string states. In
string theory, the ends of the open string states are believed to be
constrained to lie on extended solitonic objects called \Dpbranes.

Using natrual units:
\begin{equation}
  \label{eq:Kahler_function}
  G(\phi, \overline{\phi}) =
  \frac{K(\phi,\overline{\phi})}{\tilde{M}^2_{Pl}}
  + \ln
  \left(
    \frac{W(\phi)}{\tilde{M}^3_{Pl}}
  \right)
  + \ln
  \left(
    \frac{W^*(\overline{\phi})}{\tilde{M}^3_{Pl}}
  \right)
\end{equation}

$K(\phi,\overline{\phi})$ is the K\"ahler potential, a real function of
chiral superfields. This may be expaned in powers of $C_a$:
\begin{equation}
  \label{eq:Kahler_potential}
  K = 
  \overline{K}(h,\overline{h})
  + \tilde{K}_{\overline{a}{b}}(h,\overline{h}) \overline{C}_{\overline{a}} C_b
  + 
  \left[
    \frac{1}{2}Z_{ab}(h,\overline{h})C_a C_b + h.c. \right]
  + ...
\end{equation}

$\tilde{K}_{\overline{a} b}$ is the K\"ahler metric. $W(\phi)$ is the
superpotential, a holomorphic function of chiral superfields:
\begin{equation}
  \label{eq:superpotential}
  W = \hat{W}(h) + \frac{1}{2}\mu_{ab}(h)C_a C_b +
  \frac{1}{6}Y_{abc}C_a C_b C_c + ...
\end{equation}

We expect the supersymmetry to be broken; if it is broken, then the
auxilliary fields $F_\phi \ne 0$ for some $\phi$. Lacking a model of
SUSY breaking we can proceed no further without paramaterising our
ignorance. We do this using goldstino angles. We introduce a matrix,
$P$ that canonically normalises the K\"ahler metric, $P^\dagger
K_{\overline{J}I} P = 1$
\footnote{ The subscripts on the K\"ahler potential $K_I$ means
$\partial_I K$. However, the subscripts on the F-terms are just
labels. } \cite{P_Matrix:introduction}.  We also introduce a column
vector $\Theta$ which satisfies $\Theta^\dagger\Theta = 1$. We are
completely free to paramaterise $\Theta$ in any way which satisfies
this constraint.

Then the un-normalised soft terms and trilinears appear in the soft
SUGRA breaking potential \cite{Brignole:1997dp}:
\begin{equation}
  \label{eq:soft_potential}
  V_{\mathrm{soft}} = m^2_{\overline{a}b} \overline{C}_{\overline{a}} C_b
    + \left( \frac{1}{6}A_{abc} Y_{abc} C_a C_b C_c + h.c. \right) + ...
\end{equation}

The non-canonically normalised soft trilinears are then:
\begin{eqnarray}
  \nonumber
  A_{abc} Y_{abc}
  &=&
  \frac{\displaystyle \hat{W}^*}{|\displaystyle \hat{W}| }
  e^{\overline{K}/2} F_m
  \left[
    \overline{K}_m Y_{abc} + \partial_m Y_{abc} - 
    \left(
      \left(\tilde K^{-1}\right)
    \right. 
  \right. \partial_m
  \tilde K_{\overline{e}a} Y_{dbc}
  \\
  \label{eq:unnomalised_trilinears}  
  && 
  \left. 
    \left.
      { \color{white} \tilde K^-1_a }
      {} + ( a \leftrightarrow b ) 
      + ( a \leftrightarrow c )
    \right)
  \right]
\end{eqnarray}

In this equation, it should be noted that the index $m$ runs over $h,
C$. However, by definition, the hidden sector part of the K\"ahler
potential and the K\"ahler metrics are independent of the matter
fields.

Assuming that the terms $\partial_C Y_{abc} \ne 0$,
the canonically normalised equation for the trilinear is:
\begin{equation}
  \label{eq:normalised_trilinear}
  A_{abc} = F_I 
  \left[
    \overline{K}_I - \partial_I \ln
    \left(
      \tilde K_a \tilde K_b \tilde K_c
    \right)
  \right]
  + F_m \partial_m \ln Y_{abc}
\end{equation}
If the Yukawa heirarchy is taken to be generated by a Froggatt-Nielsen
field, $\phi$ such that $Y \propto \phi^p$, then we expect $F_\phi
\propto m_{3/2} \phi$, and then $F_\phi \partial_\phi \ln Y \propto
m_{3/2}$ and so even though these fields are expected to have heavily
sub-dominant F-terms, they contribute to the trilinears on an equal
footing as the moduli.

If the K\"ahler metric is diagonal and non-canonical, then the
canonically normalised scalar mass-squareds are given by
\begin{equation}
  \label{eq:normalised_scalars}
  m_a^2 = m^2_{3/2} - F_{\overline{J}} F_I \partial_{\overline{J}} \partial_I 
  \left( \ln \tilde{K_a} \right)
\end{equation}

And the gaugino masses are given by
\begin{equation}
  \label{eq:normalised_gauginos}
  M_\alpha = \frac{1}{2 \mathrm{Re} f_\alpha } F_I \partial_I f_\alpha
\end{equation}

Where $f_\alpha$ is the `gauge kinetic function'. $\alpha$ enumerates
$D$-branes in the model.  In type~I string models without twisted
moduli these have the form $f_9 = S\; ; \; f_{5^i} = T^i$.

Specifically, we use a K\"ahler potential that doesn't have any
twisted-moduli \cite{Ibanez:1998rf}:
\begin{eqnarray}
  \nonumber K &=& -\ln \left( S + \overline{S} -
  \left|C_1^{5_1}\right|^2 - \left|C_2^{5_2}\right|^2 \right) -\ln
  \left( T_1 + \overline{T}_1 - \left|C^9_1\right|^2 -
  \left|C_3^{5_3}\right|^2 \right) \\ \nonumber && {} - \ln \left( T_2
  - \overline{T}_2 - \left|C_2^9\right|^2 - \left|C_3^{5_1}\right|^2
  \right) -\ln \left( T_3 - \overline{T}_3 -\left|C_3^9\right|^2 -
  \left|C_2^{5_1}\right|^2 - \left|C_1^{5_1}\right|^2 \right) \\
  \nonumber && {} + \frac{\left|C^{5_1 5_2}\right|^2
  }{\left(S+\overline{S}\right)^{1/2}\left(T_3+\overline{T}_3\right)^{1/2}}
  + \frac{\left|C^{95_1}\right|^2}{\left(T_2 +
  \overline{T}_2\right)^{1/2}\left(T_3 + \overline{T}_3\right)^{1/2}}
  \\
  \label{eq:62}
  && {} +
  \frac{\left|C^{95_2}\right|^2}{\left(T_1+\overline{T}_1\right)^{1/2}\left(T_3+\overline{T}_3\right)^{1/2}}
\end{eqnarray}

The notation is that the field theory scalars, the dilaton $S$ and the
untwisted moduli $T_i$ originate from closed strings. Open string
states $C^b_i$ are required to have their ends localised onto
\Dbranes{}. The upper index then specifies which brane(s) their ends
are located on, and if both ends are on the same brane, the lower
index specifies which pair of compacitified extra dimensions the
string is free to vibrate in.

\section{Paramaterised trilinears for the 42241 Model}
\label{sec:param-tril-42241}

We here write the general form of the trilinear paramaters $A_{ijk}$ assuming
nothing about the form of the Yukawa matricies.
\begin{eqnarray}
  \nonumber
  A_{C_1^{5_1} C^{5_1 5_2} C^{5_1 5_2}}
  &=&
  \sqrt{3} m_{3/2}
  \left\{
    X_S
      \left[
        1 +
        \left(S + \overline{S}\right)\partial_S \ln Y_{abc} 
      \right]
    \right. 
  \\
  &
  \nonumber
  & 
  {} + X_{T_1}
    \left[
      -1 + \left(T_1 + \overline{T}_1 \right)\partial_{T_1} \ln Y_{abc}
    \right]
  \\
  &&
  \nonumber
  {} + X_{T_2}
  \left[
    -1 + \left(T_2 + \overline{T}_2 \right) \partial_{T_2} \ln Y_{abc}
  \right]
  \\
  \nonumber
  &&
  {} + X_{T_3}
    \left(T_3 + \overline{T}_3 \right)\partial_{T_3} \ln Y_{abc}
  \\
  \nonumber
  &&
  {} + X_H
    \left(S + \overline{S}\right)^{\frac{1}{2}} H \partial_H \ln Y_{abc}
    \\
  &&
  \nonumber
  {} + X_{\overline{H}}
      \left(T_3 + \overline{T}_3\right)^\frac{1}{2}\overline{H} \partial_{\overline{H}} \ln Y_{abc}
      \\
  \label{eq:trilinear_one_42241}
  &&
  \left.
    {} + X_\theta
    \left(S+\overline{S}\right)^\frac{1}{4}
    \left(T_3+\overline{T}_3\right)^\frac{1}{4}
    \theta \partial_\theta \ln Y_{abc}
  \right\}
\end{eqnarray}

\begin{eqnarray}
  \nonumber
  A_{C_1^{5_1} C_3^{5_1} C^{5_1 5_2}}
  &=&
  \sqrt{3} m_{3/2}
  \left\{
    X_S
      \left[
        \frac{1}{2} +
        \left(S + \overline{S}\right)\partial_S \ln Y_{abc} 
      \right]
    \right. 
  \\
  &&
  \nonumber
  {} + X_{T_1} 
    \left[
      -1 + \left(T_1 + \overline{T}_1 \right)\partial_{T_1} \ln Y_{abc}
    \right]
  \\
  &&
  \nonumber
  {} + X_{T_2}
    \left(T_2 + \overline{T}_2 \right) \partial_{T_2} \ln Y_{abc}
  \\
  &&
  \nonumber
  {} + X_{T_3}
    \left[
      - \frac{1}{2}
      \left(T_3 + \overline{T}_3 \right)\partial_{T_3} \ln Y_{abc}
    \right]
  \\
  \nonumber
  &&
  {} + X_H
    \left(S + \overline{S}\right)^{\frac{1}{2}}H \partial_H \ln Y_{abc}
  \\
  \nonumber
  &&
  {} +
    X_{\overline{H}}
    \left(T_3 + \overline{T}_3\right)^\frac{1}{2}\overline{H} \partial_{\overline{H}} \ln Y_{abc}
    \\
  \label{eq:trilinear_two_42241}
    &&
    \left.
    {} + X_\theta
    \left(S+\overline{S}\right)^\frac{1}{4}
    \left(T_3+\overline{T}_3\right)^\frac{1}{4}
    \theta \partial_\theta \ln Y_{abc}
  \right\}
\end{eqnarray}

\begin{eqnarray}
  \nonumber
  A_{C_1^{5_1} C_2^{5_1} C^{5_1 5_2}}
  &=&
  \sqrt{3} m_{3/2}
  \left\{
    X_S
      \left[
        \frac{1}{2} +
        \left(S + \overline{S}\right)\partial_S \ln Y_{abc} 
      \right]
    \right. 
  \\
  &&
  \nonumber
  {} + 
  X_{T_1}
    \left[
      -1 + \left(T_1 + \overline{T}_1 \right)\partial_{T_1} \ln Y_{abc}
    \right]
  \\
  &&
  \nonumber
  {} + X_{T_2} 
  \left[
    -1 + \left(T_2 + \overline{T}_2 \right) \partial_{T_2} \ln Y_{abc}
  \right]
  \\
  \nonumber
  &&
    {} +
    X_{T_3}
    \left[
      \frac{1}{2}  
      \left(T_3 + \overline{T}_3 \right)\partial_{T_3} \ln Y_{abc}
    \right]
  \\
  \nonumber
  &&
  {} + X_H
    \left(S + \overline{S}\right)^{\frac{1}{2}}H \partial_H \ln Y_{abc}
  \\
  &&
  \nonumber
    {} + X_{\overline{H}}
    \left(T_3 + \overline{T}_3\right)^\frac{1}{2}\overline{H} \partial_{\overline{H}} \ln Y_{abc}
    \\
  &&
  \label{eq:trilinear_three_42241}
  \left.
    {} + X_\theta
    \left(S+\overline{S}\right)^\frac{1}{4}
    \left(T_3+\overline{T}_3\right)^\frac{1}{4}
    \theta \partial_\theta \ln Y_{abc}
  \right\}
\end{eqnarray}

\begin{eqnarray}
  \nonumber
  A_{C_1^{5_1} C_2^{5_1} C_3^{5_1}}
  &=&
  \sqrt{3} m_{3/2}
  \left\{
    X_S
        \left(S + \overline{S}\right)\partial_S \ln Y_{abc} 
    \right. 
  \\
  &&
  \nonumber
  {} + X_{T_1}
    \left[
      -1 + \left(T_1 + \overline{T}_1 \right)\partial_{T_1} \ln Y_{abc}
    \right]
  \\
  &&
  \nonumber
  {} + X_{T_2}
    \left(T_2 + \overline{T}_2 \right) \partial_{T_2} \ln Y_{abc}
  \\
  \nonumber
  &&
  {} + X_{T_3}
    \left(T_3 + \overline{T}_3 \right)\partial_{T_3} \ln Y_{abc}
  \\
  \nonumber
  &&
  {} + X_H
    \left(S + \overline{S}\right)^{\frac{1}{2}}H \partial_H \ln Y_{abc}
  \\
  \nonumber
  &&
    {} +
    X_{\overline{H}}
    \left(T_3 + \overline{T}_3\right)^\frac{1}{2}\overline{H} \partial_{\overline{H}} \ln Y_{abc}
  \\
  \label{eq:trilinear_four_42241}
    &&
    \left.
    {} + X_\theta
    \left(S+\overline{S}\right)^\frac{1}{4}
    \left(T_3+\overline{T}_3\right)^\frac{1}{4}
    \theta \partial_\theta \ln Y_{abc}
  \right\}
\end{eqnarray}

\section{$n=1$ operators}
\label{sec:n=1-operators}

\begin{table}[htbp]
  \centering
  \framebox[14.1cm]
  {
  \begin{tabular}{|c|c||c|c|c|c|}
    \hline Operator Name & Operator Name in
    \cite{King:OperatorAnalysis} & $Q\overline{U}h_2$ &
    $Q\overline{D}h_1$ & $L \overline{E}h_1$ & $L \overline{N}h_2$ \\
    \hline $O^{Aa}$ & $O^A$ & $1$ & $1$ & $1$ & $1$ \\ $O^{Ab}$ &
    $O^B$ & $1$ & $-1$ & $-1$ & $1$ \\ $O^{Ac}$ & $O^M$ & $0$ &
    $\sqrt{2}$ & $\sqrt{2}$ & $0$ \\ $O^{Ad}$ & $O^T$ &
    $\frac{2\sqrt{2}}{5}$ & $\frac{\sqrt{2}}{5}$ &
    $\frac{\sqrt{2}}{5}$ & $\frac{2\sqrt{2}}{5}$ \\ $O^{Ae}$ & $O^V$ &
    $\sqrt{2}$ & $0$ & $0$ & $\sqrt{2}$ \\ $O^{Af}$ & $O^U$ &
    $\frac{\sqrt{2}}{5}$ & $\frac{2\sqrt{2}}{5}$ &
    $\frac{2\sqrt{2}}{5}$ & $\frac{\sqrt{2}}{5}$ \\ \rule{0mm}{4mm}
    $O^{Ba}$ & $O^C$ & $\frac{1}{\sqrt{5}}$ & $\frac{1}{\sqrt{5}}$ &
    $\frac{-3}{\sqrt{5}}$ & $\frac{-3}{\sqrt{5}}$ \\ \rule{0mm}{4mm}
    $O^{Bb}$ & $O^D$ & $\frac{1}{\sqrt{5}}$ & $\frac{-1}{\sqrt{5}}$ &
    $\frac{-3}{\sqrt{5}}$ & $\frac{3}{\sqrt{5}}$ \\ $O^{Bc}$ & $O^W$ &
    $0$ & $\sqrt{\frac{2}{5}}$ & $-3\sqrt{\frac{2}{5}}$ & $0$ \\
    $O^{Bd}$ & $O^X$ & $\frac{2\sqrt{2}}{5}$ & $\frac{\sqrt{2}}{5}$ &
    $\frac{-3\sqrt{2}}{5}$ & $\frac{-6\sqrt{2}}{5}$ \\ $O^{Be}$ &
    $O^Z$ & $\sqrt{\frac{2}{5}}$ & $0$ & $0$ & $-3\sqrt{\frac{2}{5}}$
    \\ $O^{Bf}$ & $O^Y$ & $\frac{\sqrt{2}}{5}$ & $\frac{2\sqrt{2}}{5}$
    & $\frac{-6\sqrt{2}}{5}$ & $\frac{-3\sqrt{2}}{5}$ \\ $O^{Ca}$ &
    $O^a$ & $\sqrt{2}$ & $\sqrt{2}$ & $0$ & $0$ \\ $O^{Cb}$ & $O^F$ &
    $\sqrt{2}$ & $-\sqrt{2}$ & $0$ & $0$ \\ $O^{Cc}$ & $O^E$ & $0$ &
    $2$ & $0$ & $0$ \\ $O^{Cd}$ & $O^b$ & $\frac{4}{\sqrt{5}}$ &
    $\frac{2}{\sqrt{5}}$ & $0$ & $0$ \\ $O^{Ce}$ & $O^N$ & $2$ & $0$ &
    $0$ & $0$ \\ $O^{Cf}$ & $O^c$ & $\frac{2}{\sqrt{5}}$ &
    $\frac{4}{\sqrt{5}}$ & $0$ & $0$ \\ $O^{Da}$ & $O^d$ &
    $\sqrt{\frac{2}{5}}$ & $\sqrt{\frac{2}{5}}$ &
    $2\sqrt{\frac{2}{5}}$ & $2\sqrt{\frac{2}{5}}$ \\ $O^{Db}$ & $O^e$
    & $\sqrt{\frac{2}{5}}$ & $-\sqrt{\frac{2}{5}}$ &
    $-2\sqrt{\frac{2}{5}}$ & $2\sqrt{\frac{2}{5}}$ \\ $O^{Dc}$ & $O^G$
    & $0$ & $\frac{2}{\sqrt{5}}$ & $\frac{4}{\sqrt{5}}$ & $0$ \\
    \rule{0mm}{4mm} $O^{Dd}$ & $O^H$ & $\frac{4}{5}$ & $\frac{2}{5}$ &
    $\frac{4}{5}$ & $\frac{8}{5}$ \\ \rule{0mm}{4mm} $O^{De}$ & $O^O$
    & $\frac{2}{\sqrt{5}}$ & $0$ & $0$ & $\frac{4}{\sqrt{5}}$ \\
    $O^{Df}$ & $O^f$ & $\frac{2}{5}$ & $\frac{4}{5}$ & $\frac{8}{5}$ &
    $\frac{4}{5}$ \\ $O^{Ea}$ & $O^g$ & $0$ & $0$ & $\sqrt{2}$ &
    $\sqrt{2}$ \\ $O^{Eb}$ & $O^h$ & $0$ & $0$ & $-\sqrt{2}$ &
    $\sqrt{2}$ \\ $O^{Ec}$ & $O^i$ & $0$ & $0$ & $2$ & $0$ \\ $O^{Ed}$
    & $O^j$ & $0$ & $0$ & $\frac{2}{\sqrt{5}}$ & $\frac{4}{\sqrt{5}}$
    \\ $O^{Ee}$ & $O^I$ & $0$ & $0$ & $0$ & $2$ \\ $O^{Ef}$ & $O^J$ &
    $0$ & $0$ & $\frac{4}{\sqrt{5}}$ & $\frac{2}{\sqrt{5}}$ \\
    $O^{Fa}$ & $O^P$ & $\frac{4\sqrt{2}}{5}$ & $\frac{4\sqrt{2}}{5}$ &
    $\frac{3\sqrt{2}}{5}$ & $\frac{3\sqrt{2}}{5}$ \\ $O^{Fb}$ & $O^Q$
    & $\frac{4\sqrt{2}}{5}$ & $\frac{-4\sqrt{2}}{5}$ &
    $\frac{-3\sqrt{2}}{5}$ & $\frac{3\sqrt{2}}{5}$ \\ \rule{0mm}{4mm}
    $O^{Fc}$ & $O^R$ & $0$ & $\frac{8}{5}$ & $\frac{6}{5}$ & $0$ \\
    \rule{0mm}{4mm} $O^{Fd}$ & $O^L$ & $\frac{16}{5\sqrt{5}}$ &
    $\frac{8}{5\sqrt{5}}$ & $\frac{6}{5\sqrt{5}}$ &
    $\frac{12}{5\sqrt{5}}$ \\ $O^{Fe}$ & $O^K$ & $\frac{8}{5}$ & $0$ &
    $0$ & $\frac{6}{5}$ \\ $O^{Ff}$ & $O^S$ & $\frac{8}{5\sqrt{5}}$ &
    $\frac{16}{5\sqrt{5}}$ & $\frac{12}{5\sqrt{5}}$ &
    $\frac{6}{5\sqrt{5}}$ \\ \hline
  \end{tabular}
  }
  \caption{Operator names, CGCs and names in \cite{King:OperatorAnalysis}}
  \label{tab:operator_names}
\end{table}

The $n=1$ Dirac operators are the complete set of all opearators that can be constructed from the quintilinear
$F\overline{F}h\overline{H}H$ by all possible group theoretical contractions of the indicies in
\begin{equation}
  \label{eq:fields_tensor_app}
  \mathcal{O}^{\alpha \rho y w}_{\beta \gamma x z} =
  F^{\alpha a}\overline{F}_{\beta x} h^y_a \overline{H}_{\gamma z} H^{\rho w}
\end{equation}

We define some \SU{4} invariant tensors $C$ and some \SU{2} invariant
tensors $R$ as follows
\footnote{The subscript denotes the dimension of the representation
they can create from multiplying $\mathbf{4}$ or
$\overline{\mathbf{4}}$ with $\mathbf{4}$ or
$\overline{\mathbf{4}}$. For example
$(C_{15})^{\beta\gamma}_{\alpha\rho}\overline{\mathbf{4}}_\gamma
\mathbf{4}^\rho = \mathbf{15}^\beta_\alpha$ }:
\begin{eqnarray}
  \nonumber \left(C_1\right)^\alpha_\beta &=& \delta^\alpha_\beta \\
  \nonumber \left(C_6\right)^{\rho\gamma}_{\alpha\beta} &=&
  \epsilon_{\alpha\beta\omega\chi}^{\rho\gamma\omega\chi} \\ \nonumber
  \left(C_{10}\right)^{\alpha\beta}_{\rho\gamma} &=&
  \delta^\alpha_\rho \delta^\beta_\gamma + \delta^\alpha_\gamma
  \delta^\beta_\rho \\ \nonumber
  \left(C_{15}\right)^{\beta\gamma}_{\alpha\rho} &=& \delta^\beta_\rho
  \delta^\gamma_\alpha -\frac{1}{4} \delta^\beta_\alpha
  \delta^\gamma_\rho \\ \nonumber \left(R_1\right)^x_y &=& \delta^x_y
  \\
  \label{eq:tensors}
  \left(R_3\right)^{wx}_{yz} &=& 
  \delta^x_y \delta^w_z - \frac{1}{2} \delta^x_z \delta^w_y
\end{eqnarray}

Then the six independent \SU{4} structures are:
\begin{eqnarray}
  \nonumber
  \mathrm{A}. &
  \left(C_1\right)^\beta_\alpha \left(C_1\right)^\gamma_\rho 
  &=\;\;
  \delta^\beta_\alpha \delta^\gamma_\rho
  \\
  \nonumber
  \mathrm{B}. &
  \left(C_{15}\right)^{\beta\chi}_{\alpha\sigma}
  \left(C_{15}\right)^{\gamma\sigma}_{\rho\chi}
  &=\;\;
  \delta^\beta_\rho \delta^\gamma_\alpha - 
  \frac{1}{4} \delta^\beta_\alpha \delta^\gamma_\rho
  \\
  \nonumber
  \mathrm{C}. &
  \left(C_6\right)^{\omega\chi}_{\alpha\rho}
  \left(C_6\right)^{\beta\gamma}_{\omega\chi} 
  &=\;\;
  8(\delta^\beta_\alpha \delta^\gamma_\alpha 
  - \delta^\gamma_\alpha \delta^\beta_\rho )
  \\
  \nonumber
  \mathrm{D}. &
  \left(C_{10}\right)^{\omega\chi}_{\alpha\rho}
  \left(C_{10}\right)^{\beta\gamma}_{\omega\chi} 
  &=\;\;
  2(\delta^\beta_\alpha \delta^\gamma_\rho
  + \delta^\gamma_\alpha \delta^\beta_\rho )
  \\
  \nonumber
  \mathrm{E}. &
  \left(C_1\right)^\beta_\rho
  \left(C_1\right)^\gamma_\alpha 
  &=\;\;
  \delta^\beta_\alpha \delta^\gamma_\alpha
  \\
  \label{eq:su4_structures}
  \mathrm{F}. &
  \left(C_{15}\right)^{\gamma\chi}_{\alpha\sigma}
  \left(C_{15}\right)^{\beta\sigma}_{\rho\chi}
  &=\;\;
  \delta^\gamma_\rho \delta^\alpha_\beta
  -\frac{1}{4}\delta^\gamma_\alpha \delta^\beta_\rho
\end{eqnarray}

And the six \SU{2} structures are:
\begin{eqnarray}
  \nonumber
  \mathrm{a}. &
  \left(R_1\right)^z_w
  \left(R_1\right)^x_y 
  &= \;\;
  \delta^z_w \delta^x_y
  \\
  \nonumber
  \mathrm{b}. &
  \left(R_3\right)^{zq}_{wr}
  \left(R_3\right)^{xr}_{yq} 
  &= \;\;
  \delta^x_w \delta^z_y - \frac{1}{2}\delta^x_y\delta^z_w
  \\
  \nonumber
  \mathrm{c}. &
  \epsilon^{xz}\epsilon_{yw} 
  &= \;\;
  \epsilon^{xz}\epsilon_{yw}
  \\
  \nonumber
  \mathrm{d}. &
  \epsilon_{ws}\epsilon^{xt}
  \left(R_3\right)^{sq}_{yr}
  \left(R_3\right)^{zr}_{tq} 
  &= \;\;
  \delta^x_w\delta^z_y - \frac{1}{2}\epsilon_{wy}\epsilon^{xz}
  \\
  \nonumber 
  \mathrm{e}. &
  \left(R_1\right)^z_y 
  \left(R_1\right)^x_w  
  &= \;\;
  \delta^z_y \delta^x_w
  \\
  \label{eq:su2_structurres}
  \mathrm{f}. &
  \left(R_3\right)^{zq}_{yr}
  \left(R_3\right)^{xr}_{wq} 
  &= \;\;
  \delta^x_y\delta^z_w - \frac{1}{2}\delta^x_w\delta^z_y
\end{eqnarray}

All possible $n=1$ operators were then named $O^A ... O^Z O^a...O^j$
in \cite{King:OperatorAnalysis}. We rename them here in a manner
consistent with the $n>1$ operators $O^{(n')}$, so that the names are
$O^{\Pi\pi}$ where $\Pi$ is the \SU{4} structure and $\pi$ is the
SU{2} structure.  See Table \ref{tab:operator_names} for the
translation into the names of ref.\cite{King:OperatorAnalysis} and the
CGCs.

All of these operators are operators for the case without a \UI{}
family symmetry. In the case when there is, we follow the
prescription:
\begin{equation}
  \label{eq:u1isation_of_operators}
  \mathcal{O}_{IJ}\rightarrow \mathcal{O}_{IJ}
  \left(\frac{\theta}{M_X}\right) ^{p_{_{IJ}}}
\end{equation}

Where $p_{IJ} = |X_{\mathcal{O}_{IJ}}|$ is the modulus of the charge
of the operator. If the charge of the operator is negative, then the
field $\theta$ should be replaced by the field
$\overline{\theta}$. The prescription makes the operator chargeless
under the $\UI_F$ while simultaneously not changing the dimension.

\section{$n>1$ operators}
\label{sec:n1-operators}

In the case that $n > 1$, there will be more indicies to contract,
which allows more representations, and hence more Clebsch
coefficients. To generalise the notation, it is necessary only to
construct the new tensors which create the new structures. However, it
will always be possible to contract the new indicies between the $H$
and $\overline{H}$ fields to create a singlet $H\overline{H}$ which
has a Clebsch of 1 in each sector $u,d,e,\nu$. In this case, the first
structures are the same as the old structures, but with extra $\delta$
symbols which construct the $H\overline{H}$ singlet.

Thus taking a $n=2$ operator, say $\mathcal{O}'^{Fb}$, which forms a
representation that could have been attained by a $n=1$ operator, the
Clebsch coefficients are the same. This is what we mean by
$\mathcal{O}^{n\prime\;\Pi\pi}$, as we have only used $n>1$
coefficients which are in the subset that have $n=1$ analogues.

\end{document}